# Feature-Based Compliance Control for Peg-in-Hole Assembly with Clearance or Interference Fit

Yuhang Gai, Jiuming Guo, Dan Wu, Ken Chen

*Abstract*—**This paper aims at solving mass precise peg-in-hole assembly. First, a feature space and a response space are constructed according to the relative pose and equivalent forces and moments. Then the contact states are segmented in the feature space and the segmentation boundaries are mapped into the response space. Further, a feature-based compliance control (FBCC) algorithm is proposed based on boundary mapping. In the FBCC algorithm, a direction matrix is designed to execute accurate adjustment and an integrator is applied to eliminate the residual responses. Finally, the simulations and experiments demonstrate the superiority, robustness, and generalization ability of the FBCC.**

## I. Introduction

ASSEMBLY is an indispensable step in the manufacturing process, but also a time-consuming and costly process, especially mass precise peg-in-hole assembly task. It is of great significance to improve efficiency and reduce cost through robot assembly technology. The characteristics of mass precise peg-in-hole assembly tasks are distinct. First, pegs and holes may be with clearance fit or interference fit because of the manufacturing uncertainty. Besides, there are always different nominal dimensions of pegs and holes during assembly. Therefore, the robustness of fit types and generalization ability of nominal dimensions are required.

Many researchers have worked on the automatic peg-in-hole assembly. As shown in Fig. 1, some feasible technical paths have been developed, including vision servo, force control. Moreover, combining vision servo and force control will also provide beneficial effects for peg-in-hole assembly [20].

In essence, assembly is a process of planning and controlling the relative pose of assembly objects. Consequently, it is natural and intuitive to execute the peg-in-hole assembly task under the guidance of visual servo [4]. Vision servo method has been used in aircraft assembly [1], micro assembly [2], electronic assembly [3], etc. The biggest challenge of visual servo methods is the inability to avoid assembly stress. Besides, visual servo methods are also inefficient compared with force control methods.

Force control methods can be divided into two categories: model-based force control and model-free force control.

Model-based methods are committed to establishing the mapping from the force information to the relative pose of the peg and the hole [5] [6] [7]. The basic mapping method is theoretical analysis. Some passive compliance mechanisms or active compliance control algorithms can be designed based on theoretical analysis. In a passive compliant mechanism, the robot relies on a supplemental mechanism to generate natural compliance when the end-effector tool contacts objects in the environment [5] [8] [9]. However, the compliance of passive mechanisms is always in a fixed compliance mode. Therefore, active compliance control is more widely studied and applied. The kernel of active compliance control is to control the external forces and the pose of the robot at the same time, ensuring the robot exhibits desired compliance. The contact forces and moments can be detected through force-moment sensors or state observer [14] [15]. Admittance control and impendence control are two typical methods to equip robots with active compliance [10] [11]. Mol [12] combined the admittance control and impendence control, ensuring the robot is more robust to initial misalignment. Zhang [7] applied a variable admittance control method to optimize assembly. Andre [13] obtained a haptic rendering model from CAD data to compensate for the pose deviations.

Constructing the regression model is another branch of model-based methods. Before constructing the regression model, the peg-in-hole process is always executed several times with some auxiliary methods. Then the process information is recorded for regression. Then the regression model was treated as force feedback control law [16] [17].

Model-free methods no longer attempt to establish the mapping. Reinforcement learning is a typical model-free method. Reinforcement learning method dissects a peg-in-hole task into a series of states and actions and strives to indicate the optimal action for each state [18] [19].

The biggest dilemma of the existing force control methods is that the fit type of the peg and the hole is not taken into consideration, making the methods lack robustness to manufacturing uncertainty. Besides, regression model methods and reinforcement learning methods are with distinct local attributes and hard to generalize. Once nominal dimensions of pegs and holes are changed, the methods need to be retrained. Furthermore, admittance control and impendence control are always troubled with misadjustment and residual forces and moments.

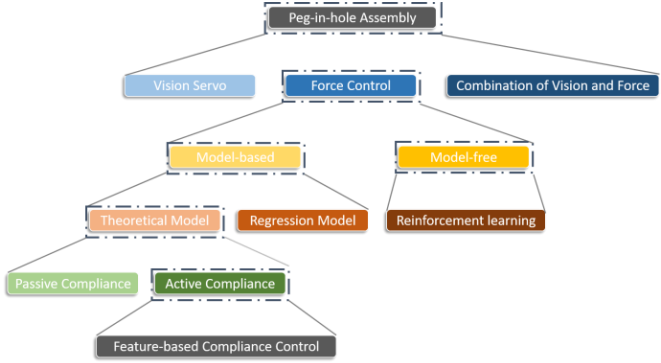

Fig. 1. Methods of robotic peg-in-hole assembly

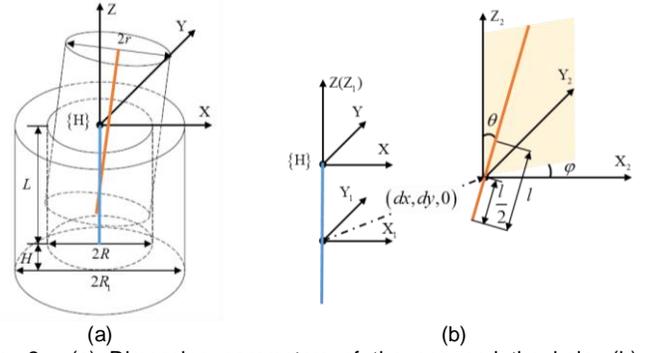

Fig. 2. (a) Dimension parameters of the peg and the hole; (b) Definitions of the extracted features.

Motivated by breaking through the dilemmas, we first construct a feature space and a response space according to the relative pose and equivalent forces and moments. Then boundary mapping is established from the response space to the feature space. Further, we propose the FBCC based on boundary mapping. Because boundary mapping will not be affected by manufacturing uncertainty and only related to the nominal dimensions of pegs and holes, the robustness and generalization ability of the FBCC can be guaranteed.

The rest of the paper is organized as follows. Section II analyzes the peg-in-hole assembly model and construct boundary mapping from the response space to the feature space. Section III develops the FBCC algorithm based on boundary mapping. The simulation verification and the experiment verification are provided in Section IV and Section V respectively.

## II. THEORETICAL ANALYSIS

In this section, our main target is to build a theoretical model to identify the contact state according to responses. First, the features of assembly are extracted. Then response calculation is conducted using the knowledge of elastic mechanics and theoretical mechanics. Subsequently, the assembly process is segmented into different states in feature space. Finally, the segmentation boundaries are mapped into response space.

### A. Feature extraction

The relative pose of cylinder pegs and holes can be determined through their axes. The degree of freedom between two line segments is five. Hence, we extract five independent features to describe the assembly process.

As shown in Fig. 2(a), a coordinate system {H} is constructed on the top surface of the hole. The origin of {H} is located at the center of the top surface and the direction of axis Z is along the axis of the hole. As shown in Fig. 2(b), the five independent features are the vertical angle $\theta$, the horizontal angle $\varphi$, the insertion depth $l$, the distance in the direction of axis X $d_x$ when the insertion depth is $l/2$, and that of axis Y $d_y$. A feature space $V$ is defined as

$$V = \{(d_x, d_y, l, \theta, \varphi) | d_x, d_y, l, \theta, \varphi \in R\} \quad (1)$$

The definition of such a feature space is not unique. To facilitate the space decomposition and state segmentation, the bases of the feature space is redefined as

$$V = \{(d_x, d_y, l, \theta_x, \theta_y) | d_x, d_y, l, \theta_x, \theta_y \in R\} \quad (2)$$

where

$$\theta_x = \tan^{-1}(\tan\theta \cos\varphi)$$
$$\theta_y = \tan^{-1}(\tan\theta \sin\varphi) \quad (3)$$

### B. Response calculation

The responses are defined as the equivalent forces and moments. The response calculation is determined through the architecture of the assembly system, especially the location of the force-moment sensor. Here we just select a common architecture where the force-moment sensor is located beneath the hole as an example. The analysis process can be easily extended to other architectures.

The hole is selected as the force analysis object. The center of the lower surface of the hole where the force-moment sensor is located is selected as the equivalent acting point. A response space is defined to integrate the forces and moments at the equivalent acting point. The response space $W$ is defined as

$$W = \{(F_x, F_y, F_f, M_x, M_y) | F_x, F_y, F_f, M_x, M_y \in R\} \quad (4)$$

where $F_x$, $F_y$, and $F_z$ are the equivalent forces in the direction of axes of {H}. $M_x$ and $M_y$ are equivalent moments in the direction of axis X and axis Y of {H}.

The equivalent forces and moments originate from the deformation of the contact area between the peg and the hole. Therefore, the contact deformation and stress of the contact area are analyzed point by point using the calculus method. As shown in Fig. 3(a), the whole contact area is divided into infinite slices along the direction of axis Z of {H}. The infinitesimal slice is termed as d$s$. The distance from the slice d$s$ to the lower surface of the peg is $s$. In the infinitesimal slice d$s$, the peg is projected as an ellipse and the hole is projected as an annulus. The center of the ellipse $O_p$ is

$$O_p = (dx - (l/2 + r\sin\theta - s/\cos\theta)\sin\theta\cos\varphi,$$
$$dy - (l/2 + r\sin\theta - s/\cos\theta)\sin\theta\sin\varphi) \quad (5)$$

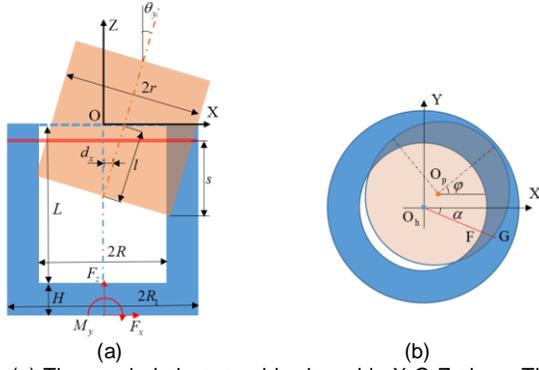

(a)　　　　　　　　　　　　　(b)

Fig. 3. (a) The peg-in-hole assembly viewed in X-O-Z plane. The red line represents the infinitesimal slice d$s$. (b) An infinitesimal slice d$s$ viewed in X-O-Y plane. The red line represents the infinitesimal sector along the direction of the angle $\alpha$.

The edges of the hole and the peg are

$$x^2 + y^2 = R^2 \quad (6)$$

$$(x - O_{px})^2 \cos^2\theta + (y - O_{py})^2 = r^2 \quad (7)$$

where $O_{px}$ and $O_{py}$ represent the coordinates of $O_p$.

As shown in Fig. 3(b), a half-line is made from the origin $O_h$ along the direction of the angle $\alpha$. The infinitesimal sector on the half-line is termed as d$\alpha$. The half-line intersects the hole at point F and the ellipse at point G. The length of the line segment $O_hF$ is constant, while that of the line segment $O_hG$ is a function of $s$ and $\alpha$. According to Lame's equation in [21], the stress of point F is

$$p = \begin{cases} k(O_hG - R)/R, & O_hG \geq R \\ 0, & O_hG < R \end{cases} \quad (8)$$

where $k$ is a constant coefficient associated with the material and dimensions of the peg and hole. Once the feature vector is given, the stress distribution in the contact area is deterministic. Therefore, the response vector can be obtained by integrating the stress distribution. Integrating the stress of each point in the contact area to the equivalent action point, the response vector is calculated as

$$F_x = \int_{s_l}^{s_u}\int_{\alpha_l}^{\alpha_u} p\cos\alpha R\,d\alpha\,ds \quad (9)$$

$$F_y = \int_{s_l}^{s_u}\int_{\alpha_l}^{\alpha_u} p\sin\alpha R\,d\alpha\,ds \quad (10)$$

$$F_f = \int_{s_l}^{s_u}\int_{\alpha_l}^{\alpha_u} \mu p R\,d\alpha\,ds \quad (11)$$

$$M_x = \int_{s_l}^{s_u}\int_{\alpha_l}^{\alpha_u} (L+H-s_u+s+\mu R)pR\sin\alpha\,d\alpha\,ds \quad (12)$$

$$M_y = \int_{s_l}^{s_u}\int_{\alpha_l}^{\alpha_u} (L+H-s_u+s+\mu R)pR\cos\alpha\,d\alpha\,ds \quad (13)$$

where $\alpha_l = -\pi$, $\alpha_u = \pi$, $s_l = 0$, $s_u = l\cos\theta$. $\mu$ is the friction coefficient between the peg and the hole. $L$ is the depth of the hole. $H$ is the thickness of the bottom layer of the hole.

*Remark* 1: The mapping constructed by force analysis will be inaccurate because of the influence of uncertainty and non-modeling factors. If the mapping is used directly as the basis of control, there will be a high likelihood of overshoot and oscillations. A more robust model will be constructed through state segmentation and boundary mapping.

*Remark* 2: The response calculation relies on the location of the force-moment sensor, which is abstracted as the equivalent action point. If the equivalent action point is changed, another similar but not the same analysis process is unavoidable. For example, if the equivalent action point is on the top surface of the peg, the definition of the $L$ and $H$ will need to be adjusted accordingly to ensure (12) and (13) is tenable.

*C. State segmentation*

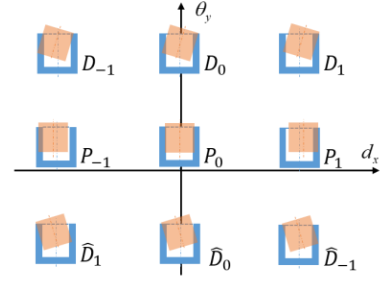

Fig. 4. The definition of contact states.

To segment the contact state of the peg and the hole in the feature space more clearly, we decompose the feature space into two subspace $V_{XOZ}$ and $V_{YOZ}$.

$$V_{XOZ} = \{(d_x, l, \theta_y) | d_x, l, \theta_y \in R\}$$
$$V_{YOZ} = \{(d_y, l, \theta_x) | d_y, l, \theta_x \in R\} \quad (14)$$

When the insertion depth $l$ is determined, we traverse the remaining two features in the subspace $V_{XOZ}$ and define nine contact states. The definition of contact states is shown in Fig. 4. The symbol $P$ represents the peg is in planar contact with the hole, while the symbol $D$ represents the peg is in double edges contact with the hole. The subscript 1 represents $d_x$ is positive, while the subscript -1 represents $d_x$ is negative. The symbol $D$ without a hat represents $\theta_y$ is positive, while the symbol $D$ with a hat represents $\theta_y$ is negative.

For a specific peg and a specific hole, the fit type is either clearance fit or interference fit. State segmentation is slightly different for the two fit types. We represent the radius of the hole as $R$ and that of the peg as $r$.

a) *Clearance fit*: When the peg and the hole are with clearance fit, $R > r$. There is always room where the peg can move and swing flexibly in the hole without contact. The boundaries of no contact are

$$\left| d_x \pm l\sin\theta_y/2 \pm r\cos\theta_y \right| = R \quad (15)$$

As the features $d_x$ and $\theta_y$ are very small numbers, the boundaries are simplified as

$$\left| d_x \pm l\theta_y/2 \pm r \right| = R \quad (16)$$

These boundaries segment the contact state $P_0$ in $V_{XOZ}$. $P_0$ is supposed to be an ideal state because there is no stress on the peg or the hole. When the peg is in double edges contact with the hole, the segmentation boundaries are $d_x = 0$ and $\theta_y = 0$.

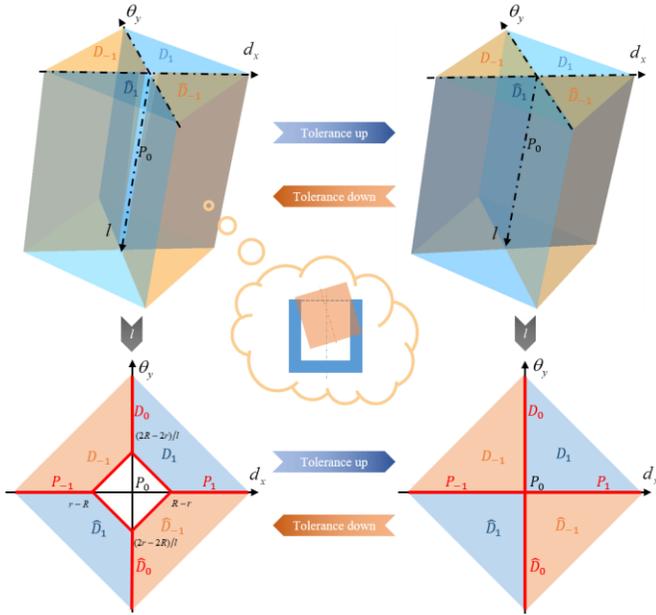

Fig. 5. The state segmentation in the feature subspace. When the peg and the hole are with clearance fit, state $P_0$ occupies a frustum of a pyramid and projects as a rhombus. As the insertion depth $l$ increases, the area of state $P_0$ continues to decrease. When the peg and the hole are with interference fit, state $P_0$ degrades to a line.

b) *Interference fit*: When the peg and the hole are with interference fit, $R < r$. The peg will always contact the hole and the stress will always exist. The segmentation boundaries of $P_0$ degrade as

$$d_x = 0, \theta_y = 0 \quad (17)$$

Because the state $P_0$ corresponds to a unique feature point, it is potential to adjust the deviations of features to zero according to the force feedback. The segmentation boundaries of other contact states are the same as clearance fit.

State segmentation in $V_{XOZ}$ is shown in Fig. 5. A similar segmentation in subspace $V_{YOZ}$ can be obtained through the same process. Then any spatial contact state in the feature space can be represented by combining the contact state in two subspaces.

*D. Boundary mapping*

Similar to the feature space, the response space is also decomposed into $W_{XOZ}$ and $W_{YOZ}$.

$$W_{XOZ} = \{(F_x, F_z, M_y) | F_x, F_z, M_y \in R\}$$
$$W_{YOZ} = \{(F_y, F_z, M_x) | F_y, F_z, M_x \in R\} \quad (18)$$

Two main segmentation boundaries in the feature subspace are $d_x = 0$ and $\theta_y = 0$. We define the integral of $\alpha$ as a function of $s$.

$$I_\alpha(s) = \int_{\alpha_l}^{\alpha_u} p \cos \alpha \, d\alpha \quad (19)$$

When $d_x = 0$, (19) is symmetric along with $s$. Hence,

$$I_\alpha(s_l + s) = -I_\alpha(s_u - s), \forall s \in [s_l, s_u], d_x = 0 \quad (20)$$

Then the value of $F_x$ is zero according to (9). The mapped boundary of $d_x = 0$ in $W_{XOZ}$ is

$$F_x = 0 \quad (21)$$

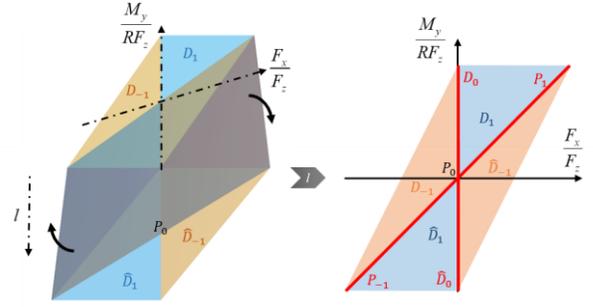

Fig. 6. The state segmentation in the response subspace. The tilted boundary rotates towards axis $F_x/F_z$ as $l$ increases.

When $\theta_y = 0$, the infinitesimal slices are the same. Hence,

$$I_\alpha(s) = \text{constant}, \forall s \in [s_l, s_u], \theta_y = 0 \quad (22)$$

Then the value of $F_x$ and $M_y$ are linearly dependent according to (9) and (13). The mapped boundary of $\theta_y = 0$ in $W_{XOZ}$ is

$$M_y / F_x = (s_l - s_u)/2 + L + H + \mu R \quad (23)$$

Based on the assumption of small features, the above boundary can be rewritten as

$$M_y / F_x = -l/2 + L + H + \mu R \quad (24)$$

$P_0$ is a special state, which can be regarded as the intersection of two boundaries. The range of state $P_0$ in feature space is differentiated according to the fit type. But the state $P_0$ in the response space is a unique point.

$$F_x = 0, M_y = 0 \quad (25)$$

State segmentation in $W_{XOZ}$ is shown in Fig. 6. We set $M_y/(RF_z)$ as the vertical axis and $F_x/F_z$ as the horizontal axis. According to (25), the state $P_0$ in response space is located at the origin. According to (21), the state $D_0$ and $\hat{D}_0$ are along the vertical axis. According to (24), the state $P_1$ and $P_{-1}$ are along the straight line through the origin. Because the contact state is continuously transformed in the feature space and the response space, the other four states $D_1$, $D_{-1}$, $\hat{D}_1$, $\hat{D}_{-1}$ in $W_{XOZ}$ are distributed between the adjacent boundaries (21) and (24). Boundary mapping from $W_{XOZ}$ to $V_{XOZ}$ is completed.

It is similar to establish boundary mapping from $W_{YOZ}$ to $V_{YOZ}$. Combining boundary mapping in X-O-Z plane and Y-O-Z plane, the spatial contact state can be judged according to the response vector.

*Remark* 1: $F_x/F_z \neq \mu$. The contact region of the peg and the hole is always a spatial surface. Therefore, contact stress at each point will not fully contribute to $F_x$, while the friction will fully contribute to $F_z$. According to (9) and (13), $F_x/F_z \leq \mu$.

*Remark* 2: Because the state $P_0$ is a point in the response space but may be an area in the feature space. The actual features may not be perceptible when the assembly process is in state $P_0$. The force controller will not offer any command when the assembly is in state $P_0$, even though the features $d_x$ and $\theta_y$ may not be zero. Therefore, the theoretical feature accuracy of force control depends on the dimension of state $P_0$ in the feature space.

Different from the mapping from a specific feature vector to a specific response vector, we only establish boundary mapping from the feature space to the response space. The accurate features are not judged according to the responses, but the contact state can be identified accurately from responses. Proposed boundary mapping is less dependent on the model information, thus making it more robust on uncertainty. Specifically, boundary mapping is consistent no matter the peg and the hole is with clearance fit or interference fit. Besides, it is highly argued that the model should be generalized easily to different nominal sizes in mass assembly tasks. Boundary mapping only depends on the nominal radius of the peg and the hole, showing its efficiency and generalization ability.

## III. CONTROL ALGORITHM

In this section, the FBCC is proposed based on boundary mapping. Then we develop a task controller framework to complete the peg-in-hole assembly task.

### A. FBCC algorithm

The main function of the controller is to supply the proper feature vector to minimize the response vector.

$$\min_{x_c} \|F_{ext}(x_c)\| \quad (26)$$

where $x_c$ represents the command feature vector. $F_{ext}$ represents the corresponding response vector.

Because of manufacturing uncertainty in mass peg-in-hole assembly, the function of $F_{ext}(x_c)$ is not definite, and the optimized feature vector is also variable. Therefore, we implement a compliance controller to realize the optimization target. The compliance control law is

$$M_d \ddot{x}_d + D_d \dot{x}_d + K_d x_d = A F_{ext}$$
$$x_c = x_{rfr} - \int x_d \quad (27)$$

where $x_{rfr}$ is the reference feature vector, $x_d$ is the adjustment feature vector. $A$ is the direction matrix related to boundary mapping. $M_d$, $D_d$, $K_d$ are the compliance parameters.

$$A = \begin{bmatrix} 1 & & & & \\ & 1 & & & \\ & & 1 & & \\ & & & -\sin\gamma & \cos\gamma/R \\ & -\sin\gamma & & & \cos\gamma/R \end{bmatrix} \quad (28)$$

where $\gamma$ is the tilt angle of the state $P_1$ and $P_{-1}$ in response subspace.

$$\gamma = \tan^{-1}(-l/2 + L + H + \mu R) \quad (29)$$

The adjustments of the feature $d_x$ and $d_y$ are different from those of $\theta_x$ and $\theta_y$. In $W_{XOZ}$, the deviation of the feature $d_x$ can be judged through the response $F_x$ independently according to boundary mapping. However, the deviation of feature $\theta_y$ must be judged through the integration of the response $F_x$ and $M_y$. The differentiation between $d_y$ and $\theta_x$ is the same in $W_{YOZ}$. The direction matrix $A$ is the reification of this differentiation in the proposed compliance control.

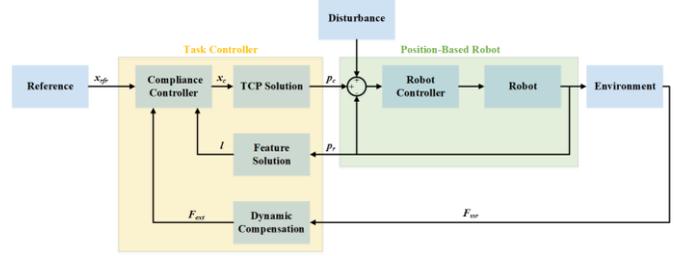

Fig. 7. The complete control framework.

---

**Algorithm 1:** Control a robot to execute assembly with the FBCC
**Input:** the reference feature vector $x_{rfr}$, the measurement data from the force-moment sensor $F_{ssr}$, the real pose of the robot $p_r$
**Output:** the command pose of the robot $p_c$
$i \leftarrow 0$  # iteration number
$x_d(i) \leftarrow 0$  # initial adjustment are always zero
$x_d(i-1) \leftarrow 0$
$F_{ext}(i) \leftarrow DynamicCompensation(F_{ssr}(i))$
$l(i) \leftarrow FeatureSolution(P_r(i))$
**for** $l(i) \leq L$ or $F_{ext}(i) \leq F_{ext\_limit}$ **do**
    $i \leftarrow i + 1$
    $F_{ext}(i) \leftarrow DynamicCompensation(F_{ssr}(i))$
    $l(i) \leftarrow FeatureSolution(P_r(i))$
    $A(i) \leftarrow CalculateDirectionMatrix(l(i))$
    $x_d(i) \leftarrow CalculateAdjustment(F_{ext}(i), A(i), x_d(i-1), x_d(i-2))$
    **if** $x_d(i) > x_{d\_limit}$ **then**
        $x_{rfr}(i) \leftarrow x_{rfr}(i-1) - x_d(i)$
    **else**
        $x_{rfr}(i) \leftarrow x_{rfr}(i-1)$
    **end if**
**end for**
$x_c(i) \leftarrow x_{rfr}(i)$
$P_c(i) \leftarrow TCPSolution(x_c(i))$

---

$$M_d = diag(M_{dp}, M_{dp}, M_{dl}, M_{do}, M_{do})$$
$$D_d = diag(D_{dp}, D_{dp}, D_{dl}, D_{do}, D_{do}) \quad (30)$$
$$K_d = diag(K_{dp}, K_{dp}, K_{dl}, K_{do}, K_{do})$$

The compliance parameters $M_d$, $D_d$, $K_d$ determine the stiffness behavior of the features and effect the adjustment speed and the control stability. In the desired assembly process, feature $l$ should increase steadily and the deviations of other features should gradually converge to zero. Therefore, we equip the features with differentiated compliance parameters. The feature $l$ is endowed with large compliance parameters $M_{dl}$, $D_{dl}$, $K_{dl}$ to ensure continuous assembly. The other features are endowed with relatively small compliance parameters to guarantee the flexibility of the assembly.

### B. Task controller framework

Because the proposed FBCC algorithm is implemented in the feature space and the response space, additional modules are necessary to connect the robot and the force-moment sensor. The complete control framework is shown in Fig. 7. The task controller includes four modules, which need to be designed. The compliance controller has been introduced in detail. The dynamic compensation module removes the dynamic forces from the measurement data of the force-moment sensor $F_{ssr}$. The mission of the TCP solution module is to translate the command feature vector $x_c$ to the command pose of the position-based robot $p_c$. The feature solution module computes

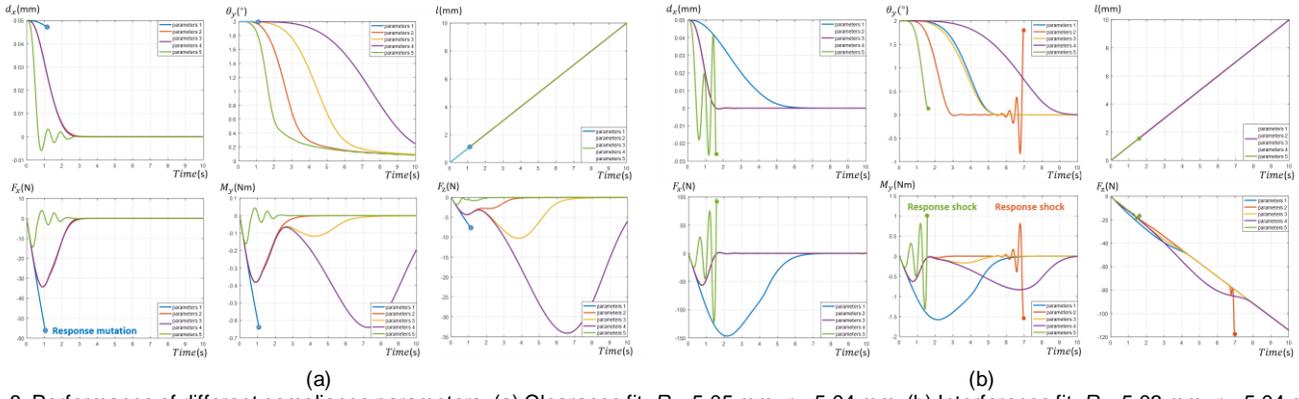
Fig. 8. Performance of different compliance parameters. (a) Clearance fit, $R$ = 5.05 mm, $r$ = 5.04 mm. (b) Interference fit, $R$ = 5.03 mm, $r$ = 5.04 mm.

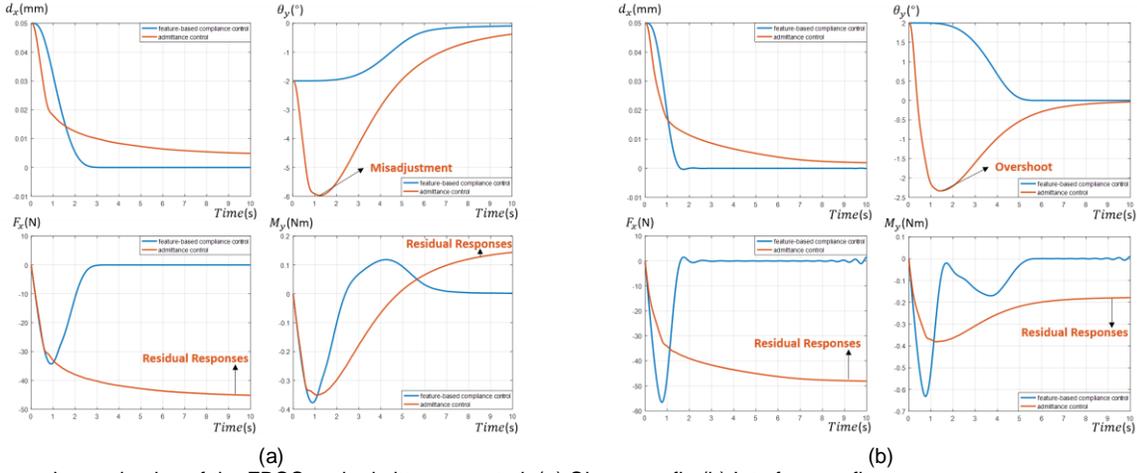
Fig. 9. Comparative evaluation of the FBCC and admittance control. (a) Clearance fit. (b) Interference fit.

the feature $l$ according to the real pose of the robot $p_r$. The complete FBCC algorithm is shown as algorithm 1.

## IV. SIMULATION

The simulation is implemented in MATLAB Simulink software. A custom module used to simulate contact stress and responses is constructed by finite element method.

### A. Exploration of proper compliance parameters

The proper parameters must keep feature trajectories and the response trajectories stable enough and reach the steady state as quickly as possible. Besides, the maximum value of the response trajectories should be limited.

Five typical compliance parameters were selected in Table I. The initial deviation of feature $d_x$ and $d_y$ is set to 0.05 mm and that of $\theta_x$ and $\theta_y$ is 2°. The final insertion depth $L$ is 10 mm. The simulation is implemented under the condition of clearance fit and interference fit separately.

As the feature trajectories and the response trajectories are the same in two feature subspaces and two response subspaces, only the trajectories in $V_{XOZ}$ and $W_{XOZ}$ are shown in Fig. 8. When using compliance parameters 1, a non-ideal huge response mutation occurs, which may terminate the assembly. When applying parameters 2 and 5, the response shock also damages the assembly.

Summing up the above observations, excessive compliance parameters delay the adjustment, which ultimately leads to great responses, while too small compliance parameters will damage the stability. Hence, parameters 3 are considered to be proper and is applied in the subsequent simulation.

TABLE I
THE COMPLIANCE PARAMETERS IN SIMULATIONS

| Label | $M_{dp}$ | $K_{dp}$ | $M_{do}$ | $K_{do}$ | $M_{dl}$ | $K_{dl}$ | $D_d$ |
|---|---|---|---|---|---|---|---|
| 1 | $10^5$ | $10^7$ | $10^1$ | $10^3$ | $10^{10}$ | $10^{12}$ | |
| 2 | $10^4$ | $10^6$ | $10^0$ | $10^2$ | $10^{10}$ | $10^{12}$ | |
| 3 | $10^4$ | $10^6$ | $10^1$ | $10^3$ | $10^{10}$ | $10^{12}$ | $\sqrt{M_d K_d}$ |
| 4 | $10^4$ | $10^6$ | $10^2$ | $10^4$ | $10^{10}$ | $10^{12}$ | |
| 5 | $10^3$ | $10^5$ | $10^{-1}$ | $10^1$ | $10^{10}$ | $10^{12}$ | |

### B. Comparative evaluation with admittance control

The proposed FBCC is compared with admittance control here to verify its effectiveness. The compliance parameters, the dimensions of the peg and the hole, and the final insertion depth $L$ are inherited. The initial deviation of feature $d_x$ is 0.05 mm and that of $\theta_y$ is -2° or 2°. The feature trajectories and the response trajectories in $V_{XOZ}$ and $W_{XOZ}$ are plotted in Fig. 9.

At the beginning of the assembly, the response $F_x$ and $M_y$ increase rapidly. Predictably, the response $F_x$ and $M_y$ will continue to increase until jamming occurs if no force feedback is implemented. Two control laws can both eliminate the initial deviations. With the guidance of the FBCC or admittance control, the responses begin to decline after increasing to a certain value. Finally, the feature $d_x$ and $\theta_y$ are gradually reaching a stable value.

Compared with admittance control, the FBCC has made two

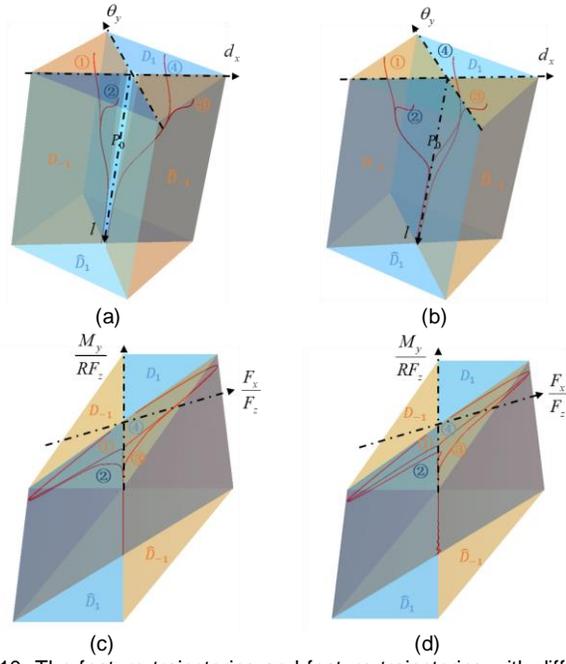

(a) (b)

(c) (d)

Fig. 10. The feature trajectories and feature trajectories with different initial deviations of features. (a) Clearance fit. Feature space. (b) Interference fit. Feature space. (c) Clearance fit. Clearance fit. (d) Interference fit. Response subspace. The initial deviations: ① $d_x = -0.05mm, \theta_y = 2°$. ② $d_x = -0.1mm, \theta_y = -1°$. ③ $d_x = 0.1mm, \theta_y = -1°$. ④ $d_x = 0.05mm, \theta_y = 2°$.

main breakthroughs in rapidity and accuracy, especially for orientation features.

1) Admittance control always causes a misadjustment or overshoot. In Fig. 9(a), the deviations of features $d_x$ and $\theta_y$ are of different sign. The feature $\theta_y$ gets disorientated to -6°. In Fig. 9(b), the deviations of features $d_x$ and $\theta_y$ are of the same sign. The adjustment of the feature $\theta_y$ is in the right direction but a large overshoot of 2.5° appears. The misadjustment or overshoot is devastating for precise assembly. The proposed FBCC solves the problems mentioned above well.

2) Because admittance control ignores historical information, the residual responses have to exist to supply adjustments, which means that the residual responses cannot be eliminated. By utilizing an integrator in the FBCC, historical adjustments are summed as the current command feature vector and the residual responses are reduced to zero.

## C. Verification of robustness and generalization ability

To further verify the robustness and generalization ability of the FBCC, some different initial deviations of features are chosen. The nominal radius is reset to be 10 mm. The tolerance and the compliance parameters are inherited. The feature trajectories and the response trajectories are plotted in Fig. 10.

As shown in Fig. 10(a) and Fig. 10(b), four lines represent the feature trajectories and response trajectories when the peg and the hole are with clearance fit. All the feature trajectories tend to transfer into the state $P_0$ no matter which initial state they belong to. The transformation process is almost completed in the initial state, and hardly into other states except $P_0$, which is consistent with the fact that there is no overshoot when implementing the FBCC. The response trajectories also transfer towards the state $P_0$ after their initial increase.

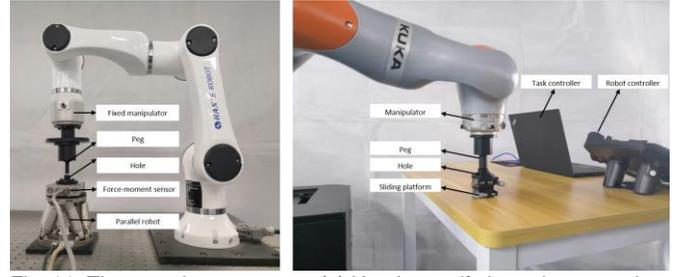

Fig. 11. The experiment system. (a) Used to verify boundary mapping. (b) Used to execute the peg-in-hole assembly.

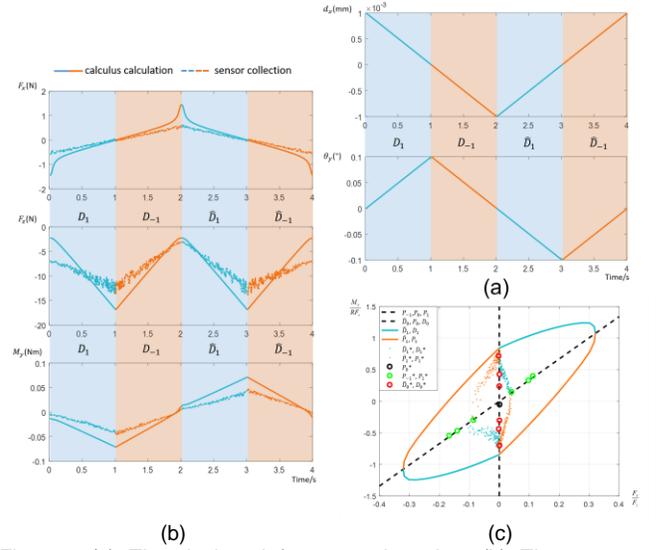

(a)

(b) (c)

Fig. 12. (a) The designed feature trajectories. (b) The response trajectories and response calculation results. (c) The response trajectories and response calculation results mapped in state segmentation of response subspace.

As shown in Fig. 10(c) and Fig. 10(d), the transformation of feature trajectories and response trajectories are similar when the fit type is interference fit. The FBCC can still eliminate these different initial deviations and transfer the trajectories into state $P_0$ successfully. Because the control requirements are much higher when the peg and the hole are with interference fit, the transformation process is not as smooth as that of clearance fit.

From the above results, the FBCC is declared as a robust algorithm for the initial deviations and fit types. In practical application, the FBCC can be taken as the next step of position control to expand the task ability of robots. The maximum initial deviations are ±0.05 mm and ±2°, which are larger than the repeatable positioning accuracy of most robots. From the perspective of practical application, the robustness of the FBCC satisfies the requirements. The FBCC is also easy to generalize because only the nominal radius needs to be updated in the controller.

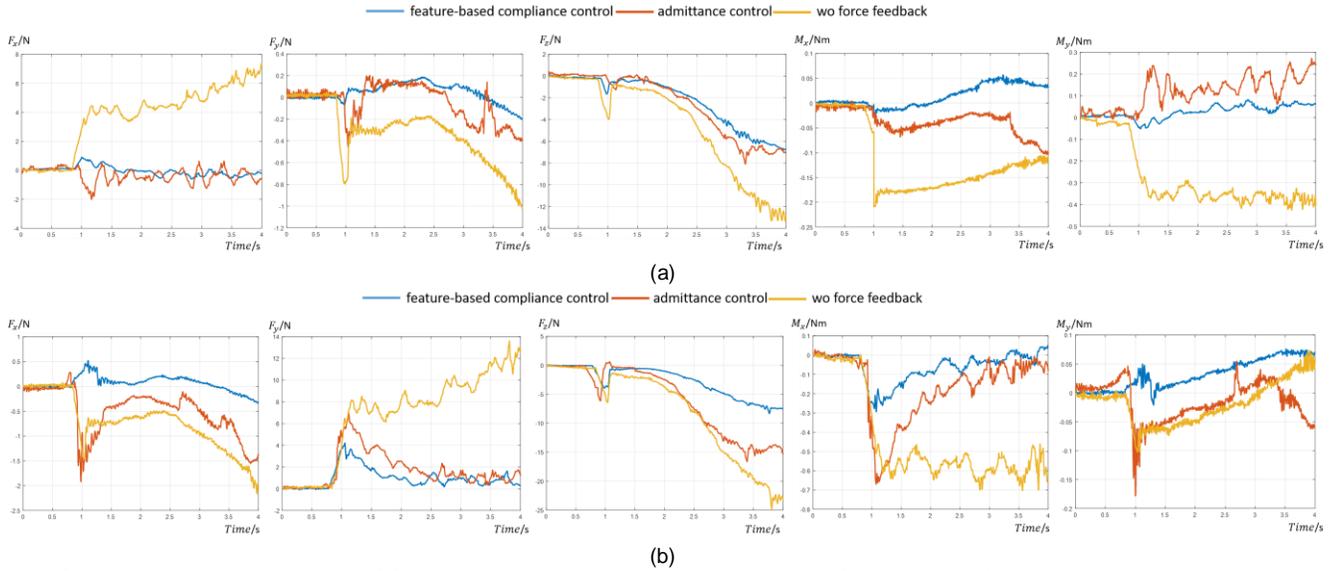

Fig. 13. Comparative evaluation of the FBCC and admittance control in the experiment. (a) Clearance fit. (b) Interference fit.

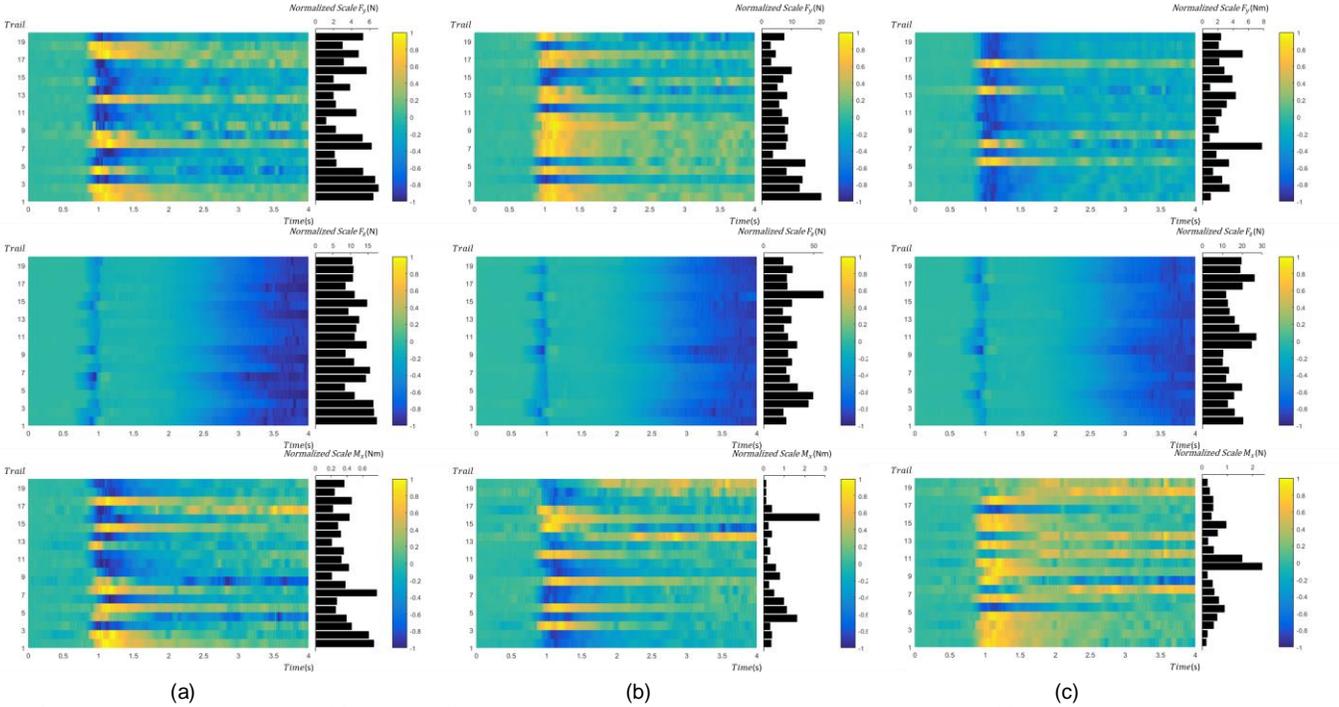

Fig. 14. The response trajectories in 20 trails. (a) Clearance fit. (b) Interference fit. (c) The nominal radius is 10 mm.

TABLE II
THE DIMENSIONS OF THE PEG AND THE HOLE

| Object | Measured diameter (mm) | | | | | Average |
|---|---|---|---|---|---|---|
| Peg | 10.037 | 10.031 | 10.041 | 10.035 | 10.029 | 10.034 |
| Hole | 10.033 | 10.031 | 10.027 | 10.032 | 10.031 | 10.031 |
| Peg | 10.113 | 10.056 | 10.047 | 10.072 | 10.097 | 10.077 |
| Hole | 10.149 | 10.135 | 10.145 | 10.155 | 10.156 | 10.148 |
| Peg | 10.044 | 10.050 | 10.043 | 10.050 | 10.044 | 10.0462 |
| Hole | 10.021 | 10.019 | 10.018 | 10.011 | 10.018 | 10.0174 |

## V. EXPERIMENT

In this section, we further conduct experiments to prove the effectiveness of theoretical analysis and control algorithm. The target of experiments is to confirm the correctness of boundary mapping as well as demonstrate the performance of the FBCC. As shown in Fig. 11, two different experiment platforms are constructed for the targets.

### A. Verification of boundary mapping

The architecture of the experiment system is shown in Fig. 11 (a). To verify the correctness of boundary mapping, it is crucial to supply accurate motion and collect precise measurement data. Hence, we arrange a parallel robot PI H-811.I2 as the motion mechanism and a force-moment sensor Kistler 9256C2 as the collection device. The repeated positioning accuracy of the parallel robot reaches ±0.06 μm and ±2 μrad, which guarantees an accurate motion. Because the force-moment sensor lacks the ability to measure $M_y$, the experiment is executed in $V_{YOZ}$ and $W_{YOZ}$. The collection threshold of the force-moment sensor reaches 2 mN, which ensures the measurement data are reliable enough. The peg is fixed on a locked manipulator. The hole is fixed on the force-moment sensor and the force-moment sensor is on the parallel robot. The material of the peg and the hole are ABS. The dimensions of the peg and the hole are shown as first group in Table II.

Because the accurate feature trajectories are supplied by the parallel robot, a calibration is required to indicate the relationship between the pose of the parallel robot and the feature vector. The core of the calibration is to find the pose of the robot where the peg and the hole are in state $P_0$. The basic assumption is that the feature $d_x = 0$ and $\theta_y = 0$ always generate the response $F_x = 0$ and $M_y = 0$. Hence, it is convenient as well as intuitive to calibrate with the guidance of the force-moment sensor. We first command the parallel robot to make the hole inserted by the peg manually until the feature $l$ comes to 5 mm. Then the motion of the parallel robot is adjusted manually to change the feature $d_x$ and $\theta_y$ until the response $F_x = 0$ and $M_y = 0$. Here the assembly is considered to enter state $P_0$ and the calibration has been accomplished.

As shown in Fig. 12(a), the trajectories of feature $d_x$ and $\theta_y$ are designed to go through all contact states. Based on the calibration, the parallel robot is commanded to execute the feature trajectories. Meanwhile, the response trajectories are collected by the force-moment sensor, which is shown in Fig. 12(b). To display the experimental results more clearly, we plot the response trajectories and response calculation results in Fig. 12(c). Due to manufacturing uncertainty of the peg and the hole, the measurement data of the force-moment sensor are quite different from response calculation results, which indicates that it is unreliable to calculate how much to adjust the feature directly according to theoretical calculation. However, boundary mapping can supply a reliable state judgment according to the response vector. Hence, boundary mapping, the foundation of the FBCC, is regarded to be stable and solid.

### B. Comparative evaluation with admittance control

The experimental system is shown in Fig. 11(b). The assembly task is completed by the manipulator KUKA LBR iiwa. The manipulator is equipped with torque sensors at each joint. Therefore, it is possible to construct a virtual force-moment sensor at the end of the manipulator. The peg is fixed on the manipulator and the hole is fixed on the table through a sliding platform. The sliding platform can be adjusted manually to change the initial deviations between the peg and the hole. The material of the peg and the hole are still ABS. The final insertion depth $L$ is 10 mm. Two groups of the peg and the hole with different fit types are selected in the experiment. The dimensions of pegs and holes are listed as the second group and third group in Table II.

Because the responses are detected through the virtual force-moment sensor on the top surface of the peg, the equivalent action point has to be changed and the value of $L$ and $H$ in (29) have to be also corrected. Besides, the adaption of the dynamic compensation module also needs to be modified.

The response trajectories of different control methods are shown in Fig. 13. We first focus on the response trajectories when there is no force feedback. There is a clear trend that the responses tend to be much larger if force feedback does not work, which is consistent with theoretical analysis. Both admittance control and the FBCC can improve the response trajectories. As indicated in the simulation, the admittance control is still troubled with the challenges of misadjustment or overshoot and residual responses. Besides, admittance control is more likely to cause response mutation and response shock. the FBCC solves these challenges effectively. Moreover, the FBCC always produces a smaller response $F_z$, showing its superiority.

### C. Verification of robustness and generalization ability

To demonstrate the robustness of the proposed control law against the initial deviations, we adjust the initial position of the hole through the sliding platform. The initial deviations are chosen randomly within $\pm 0.1$ mm. 20 trails are implemented for the peg and the hole with different fit types.

The response trajectories in $W_{YOZ}$ are gathered together in Fig. 14(a) and Fig. 14(b). The responses are normalized and plotted on a color map. The normalized scales of the responses are also attached as a histogram after the color map. The contact occurs about one second after the beginning of the assembly. Consequently, large responses appear at the same time. The responses continue getting larger in a few tenths of a second later. Because the initial deviations are random, the corresponding response $F_y$ and $M_x$ can be negative or positive. Then the response $F_y$ and $M_x$ gradually approach zero with the guidance of the FBCC. The response $F_z$ is always positive. When the peg and the hole are with clearance fit, the largest response $F_z$ can be constrained to around 15 N. When the peg and the hole are with interference fit, the response $F_z$ is unavoidable because the contact stress always exists. Even though, the response $F_z$ of most trails can be still optimized to less than 50 N. The whole assembly process lasts 3 s. The success rate of all trails is 100%. The results declare that boundary mapping is an intrinsic rule for peg-in-hole assembly and the FBCC is a high-level policy, which is robust to the initial deviations and manufacturing uncertainty.

Moreover, to test the generalization ability of the control strategy, we select the peg and the hole with a radius of 10 mm. We change nothing except the information of the nominal radius in the controller. 20 trails are implemented on the new peg and hole to evaluate the control effect. As shown in Fig. 14(c), the response $F_y$ and $M_x$ gradually stabilize towards zero after the initial contact. Though the final responses of some trails may not be close to zero in the color map, the normalized scales are small enough, thus making the actual responses have a limited impact on the assembly. The results declare that the FBCC is easy to generalize to different dimensions, which is significant to improve assembly efficiency.

## VI. CONCLUSION

In this paper, the assembly task is abstracted as a series of features and responses. With the guidance of state segmentation and boundary mapping, the contact state can be identified according to the responses. Then the FBCC algorithm is proposed based on boundary mapping. Further, the complete task control framework is accomplished to execute peg-in-hole assembly. Compared with admittance control, the FBCC guarantees more accurate adjustments and less residual responses. Besides, the robustness of the FBCC to initial deviations and fit types is verified, which ensures the realization of mass precise peg-in-hole assembly. Moreover, the generalization ability of the FBCC is also demonstrated, which is significant for assembly efficiency.

However, boundary mapping, the basic of the FBCC, is constructed with the assumption that the peg and the hole are columniform. In the future, boundary mapping for pegs and holes of other shapes should be explored. Besides, the mapping and control algorithm combined with machine learning or reinforcement learning is another direction of research.